\documentclass[aps,prl,a4paper,reprint,superscriptaddress,showpacs,amsfonts,amssymb]{revtex4-1}%

\usepackage{graphicx}
\usepackage{dcolumn}
\usepackage{bm}

\begin{document}

\title{Momentum-Resolved Ultrafast Electron Dynamics in Superconducting Bi$_{2}$Sr$_{2}$CaCu$_{2}$O$_{8+\delta}$}

\author{R. Cort\'{e}s}
\affiliation{Fachbereich Physik, Freie Universit\"{a}t Berlin, Arnimallee 14, D-14195 Berlin, Germany} 
\affiliation{Abt. Physikalische Chemie, Fritz-Haber-Institut d. MPG, Faradayweg 4-6, D-14195 Berlin, Germany}

\author{L. Rettig}
\affiliation{Fachbereich Physik, Freie Universit\"{a}t Berlin, Arnimallee 14, D-14195 Berlin, Germany} 
\affiliation{Fakult\"{a}t f\"{u}r Physik, Universit\"{a}t Duisburg-Essen, Lotharstr. 1, D-47048 Duisburg, Germany}

\author{Y. Yoshida}
\affiliation{National Institute of Advanced Industrial Science and Technology, Tsukuba, Ibaraki 305-8568, Japan}

\author{H. Eisaki}
\affiliation{National Institute of Advanced Industrial Science and Technology, Tsukuba, Ibaraki 305-8568, Japan}

\author{M. Wolf}
\affiliation{Fachbereich Physik, Freie Universit\"{a}t Berlin, Arnimallee 14, D-14195 Berlin, Germany} 
\affiliation{Abt. Physikalische Chemie, Fritz-Haber-Institut d. MPG, Faradayweg 4-6, D-14195 Berlin, Germany}

\author{U. Bovensiepen}
\affiliation{Fakult\"{a}t f\"{u}r Physik, Universit\"{a}t Duisburg-Essen, Lotharstr. 1, D-47048 Duisburg, Germany}

\date{\today}

\begin{abstract}
The non-equilibrium state of the high-$T_{\mathrm{c}}$ superconductor Bi$_{2}$Sr$_{2}$CaCu$_{2}$O$_{8+\delta}$ and its ultrafast dynamics have been investigated by femtosecond time- and angle-resolved photoemission spectroscopy well below the critical temperature. We probe optically excited quasiparticles at different electron momenta along the Fermi surface and detect metastable quasiparticles near the antinode. Their decay through e-e scattering is blocked by a phase space restricted to the nodal region. The lack of momentum dependence in the decay rates is in agreement with relaxation dominated by Cooper pair recombination in a boson bottleneck limit. 
\end{abstract}


\maketitle

The pairing mechanism responsible for the high$-T_{c}$ superconductivity in the cuprates remains one of the most challenging problems of current Solid State Physics, after more than two decades of research. In this context, angle-resolved photoemission spectroscopy (ARPES) has proven to be a very powerful experimental technique, providing information on the single-particle spectral function of these materials with a very high energy and momentum resolution \cite{Damascelli03}. However, it gives limited information on the coupling of single particle states with collective excitations, which seems to be essential to understand the ground state of the high$-T_{c}$ superconductors (HTSC) \cite{Damascelli03,Kordyuk06}. Additional information can be obtained from femtosecond (fs) time-resolved optical and THz techniques \cite{Demsar99,Gedik04,Kusar05,Kaindl05,Gedik05,Kusar08,Liu08,Giannetti09}, which allow to study the quasiparticle (QP) interactions responsible for the relaxation of a photoexcited non-equilibrium state. This, in the case of the HTSC, is considered to provide a tool to study the interactions responsible for the pairing of QPs forming Cooper pairs. The analysis of QP decay dynamics has led to a controversy whether the decay follows a bimolecular recombination or proceeds in a boson bottleneck regime~\cite{Kaindl05,Gedik05,Kabanov05,Kusar08}. However, these experiments inherently lack momentum resolution and thus can only be related to the electronic band structure in an indirect way. Finally, theoretical works related with these optical studies \cite{Howell04,Gedik04} have provided further insight into the QP dynamics, but they have arrived to conclusions about the metastability of the nonequilibrium QPs which were up to now difficult to prove by experimental means.
 
Complementary to ARPES and all optical time-resolved techniques, femtosecond time- and angle-resolved photoemission spectroscopy (trARPES) provides momentum and energy resolved information on the single particle spectral function and its temporal evolution, allowing a direct investigation of the QP relaxation along the Fermi surface (FS). However, first investigations on cuprate superconductors using trARPES \cite{Perfetti07} did not develop a fully momentum-resolved study of their ultrafast dynamics nor used low enough excitation densities to avoid the instant vaporization of the superconducting condensate.

In this letter, we report on the ultrafast electron dynamics in superconducting Bi$_{2}$Sr$_{2}$CaCu$_{2}$O$_{8+\delta}$ (Bi-2212) investigated at different points of its normal state FS by trARPES. Our data show that the density of nonequilibrium QPs created by photoinduced breaking of Cooper pairs is momentum dependent and related to the size of the superconducting gap. In contrast, the recombination rate of these QPs shows no sign of momentum or excitation density dependence. Our results provide experimental evidence of the transient stabilization of QPs off the node, due to scattering phase space restrictions caused by energy and momentum conservation in a \textit{d}-wave superconductor. They also demonstrate that the net QP recombination rate in Bi-2212 is determined by the decay rate of the bosons emitted in this process (boson bottleneck).

\begin{figure}
\includegraphics{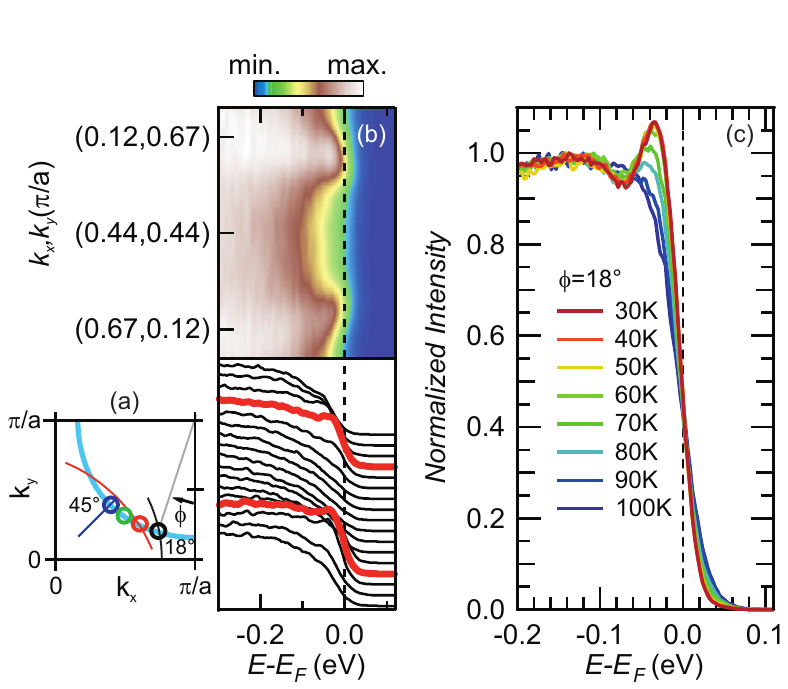}
\caption{(Color online) (a) Sketch of the normal state FS of Bi-2212. Circles mark the FS angles, $\phi=18^{\circ}, 27^{\circ}, 37^{\circ}$ and $45^{\circ}$, considered in this work. Some of the arcs cutting the FS along which the ARPES spectra were taken are also shown. (b) ARPES spectra and their representation as a false color plot, measured along the red arc in (a). The spectra measured at the FS ($\phi=27^{\circ}$ and $63^{\circ}$) are highlighted with red thick lines in the lower panel. (c) ARPES spectra measured at $\phi=18^{\circ}$ as a function of the temperature. 
\label{fig_ARPES}}
\end{figure}

The Bi-2212 samples studied in this work were nearly optimally doped single crystals with a transition temperature $T_\mathrm{c}=88$~K. They were cleaved in situ in ultrahigh vacuum ($\sim 8\cdot 10^{-11}$~mbar) at 30 K, where the experiments were carried out. In the trARPES measurements the samples were excited by 55 fs laser pulses with a photon energy of 1.5 eV (pump beam), at 300 kHz repetition rate. Absorbed fluences, $F$, between 6 and 139~$\mu$J/cm$^{2}$ were used. The transient electron distribution was probed by time-delayed 80 fs, 6 eV  laser pulses (probe beam) giving rise
to photoelectrons, which were detected by a time-of-flight spectrometer. The energy resolution was typically 50 meV, the momentum resolution was 0.05~\AA$^{-1}$ and the time resolution $<100$~fs, see \cite{Schmitt08} for details. By means of a slanted sample holder \cite{Schmitt08} it was possible to reach points along the FS corresponding to FS angles, $\phi$, between $45^{\circ}$ (nodal point) and $18^{\circ}$ (Fig.\ref{fig_ARPES}(a)), in spite of the low photon energy of the probe beam.

The equilibrium electronic band structure around the Fermi level was studied by laser-based ARPES, using only the 6~eV beam. ARPES spectra (Fig.\ref{fig_ARPES}(b)) were taken along arcs in the reciprocal space cutting the normal state FS (Fig.\ref{fig_ARPES}(a)). In the spectra measured at the FS, a sharp peak is observed, which is known to be a direct consequence of the superconducting state \cite{Damascelli03}. The temperature dependence of this superconducting peak is shown for $\phi=18^{\circ}$ in Fig.\ref{fig_ARPES}(c). Its  disappearance above $T_{c}$ corroborates its relation with superconductivity.

\begin{figure}
 \includegraphics{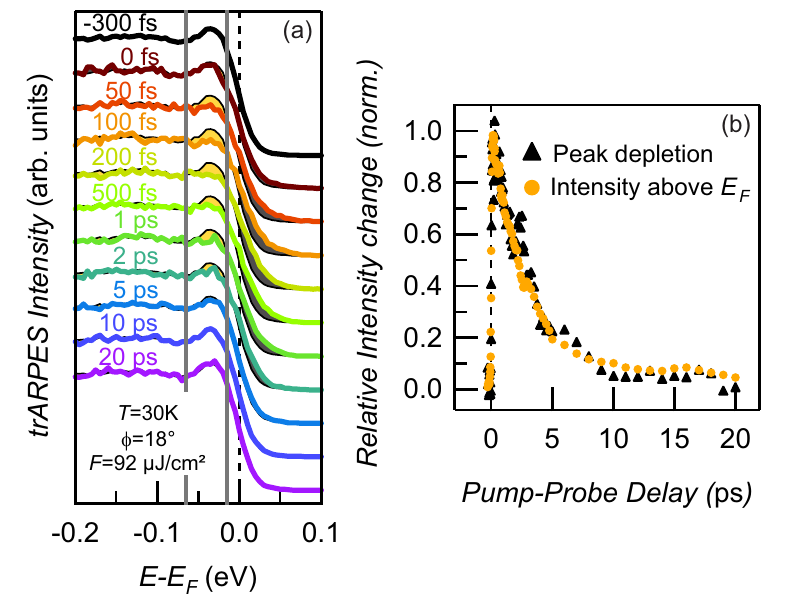}
 \caption{(Color online) (a) trARPES spectra measured at several pump-probe delays. The depletion of the superconducting peak and the increase of spectral weight above the Fermi level, in relation to the spectrum measured before optical excitation, are shadowed in yellow (bright) and gray (dark), respectively. (b) Depletion of the superconducting peak (yellow (bright) area between the vertical gray lines in (a)) and increase of the intensity above the Fermi level (gray (dark) area in (a)) as a function of the pump-probe delay. \label{fig_trSCpeak}}
 \end{figure}

The excitation of the sample by the 1.5 eV pump pulse produces a depletion of the superconducting peak, as well as an increase of the spectral weight above the Fermi level, $E_{F}$, (Fig.\ref{fig_trSCpeak}) which are different than the ones produced by a mere increase of the temperature, see Fig.\ref{fig_ARPES}(c). We find that both quantities have the same evolution with the pump-probe delay. Thereby we show that the increase of the spectral weight at $E>E_{F}$ corresponds to the creation of a nonequilibrium density of QPs by breaking Cooper pairs and the decrease of that spectral weight can be attributed to the recombination of these QPs. At this stage we conclude that the time dependent spectral weight above $E_{F}$ directly reflects the dynamics of the recovery of the superconducting condensate after photoexcitation. We proceed now by a momentum dependent analysis of the evolution of that spectral weight at $E>E_{F}$.

\begin{figure}
\includegraphics{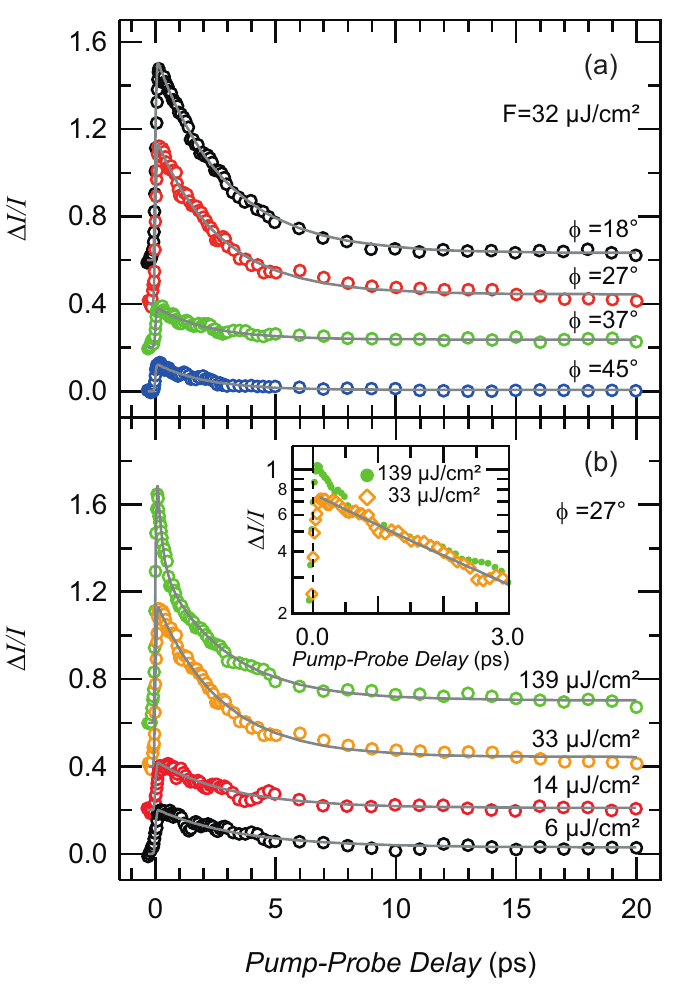}
\caption{(Color online) Relative trARPES intensity change above the Fermi level, $\Delta$$I/I$, measured at different FS angles, $\phi$, using a pump fluence $F=32$~$\mu$J/cm$^{2}$ (a) and measured at $\phi=27^{\circ}$, using different pump fluences (b). A zoom of the spectra measured at $\phi=27^{\circ}$ with $F=139$~$\mu$J/cm$^{2}$ and $F=33$~$\mu$J/cm$^{2}$, using a logarithmic vertical scale, is shown as an inset in (b). The fitting to exponential decays (see text) is shown as thin gray lines. In the inset, only the fitting of the spectrum measured at 33~$\mu$J/cm$^{2}$ to a single-component exponential decay function for t$>100$~fs is shown. To fit the spectrum measured at 139~$\mu$J/cm$^{2}$ an additional component is needed.\label{fig_XCs}}
 \end{figure}

trARPES spectra were measured at four points of the FS, with $\phi=18^{\circ}, 27^{\circ}, 37^{\circ}$ and $45^{\circ}$ (Fig.\ref{fig_ARPES}(a)), for different pump fluences, $F=6, 14, 33, 139$~$\mu$J/cm$^{2}$. Next, the normalized trARPES intensity change with respect to the intensity before the arrival of the pump pulse, $\Delta$$I(t)/I$, was determined for $E>E_{F}$ (Fig.\ref{fig_XCs}). The decay of $\Delta$$I(t)/I$ in the measurements with $F\leq 33$~$\mu$J/cm$^{2}$ was fitted to a single-component exponential decay, $\Delta$$I(t)/I=A \exp(-t/\tau)+B$, convoluted with a 100 fs width Gaussian accounting for the time resolution. $A$ is the excitation amplitude, $\tau$ is the relaxation time of the nonequilibrium QPs and $B$ accounts for heat diffusion effects \cite{Schoenlein87}. For larger fluences, an additional decay component with smaller $\tau$ is observed in $\Delta$$I(t)/I$, see the inset of Fig.\ref{fig_XCs}(b). We fit these data by a biexponential decay, accounting for the slow and fast component, see Fig.\ref{fig_XCs}.

\begin{figure}
\includegraphics{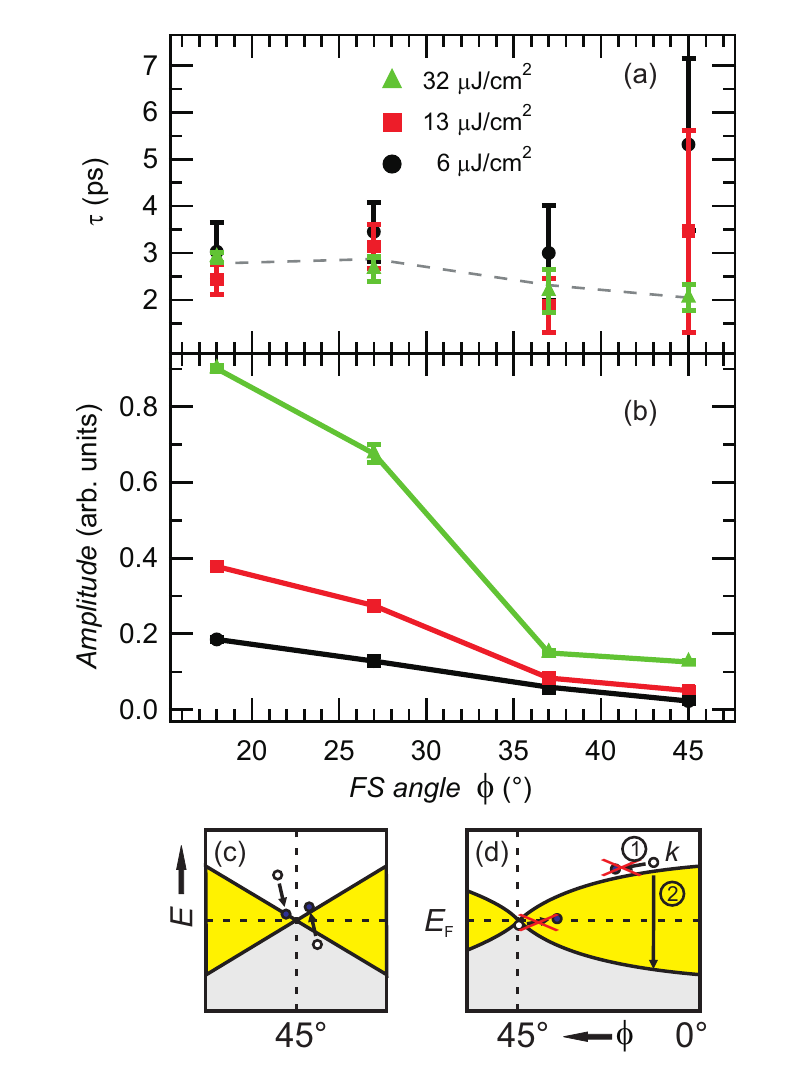}
\caption{(Color online) Relaxation time (a) and excitation amplitude (b), obtained from the fitting to exponential decays (see text), as a function of the FS angle, $\phi$. Lines are guides to the eye. Panels (c,d) illustrate potential decay processes of photoexcited quasiparticles discussed in the text. The energy per pump pulse was kept constant, which lead in angle-dependent studies to a variation in $F$ due to changes in reflectivity and illuminated area. The data were corrected accordingly.\label{fig_tauAndA}}
 \end{figure}

First we focus on the slower contribution and its momentum dependence. Fig.~\ref{fig_tauAndA} shows the momentum-dependent amplitudes $A$ and decay times $\tau$ obtained by fitting $\Delta$$I(t)/I$. All the fluences considered here show -- within error bars -- a constant $\tau\sim2.5$~ps, see panel (a), and a decrease in $A$ with increasing $\phi$, panel (b). Albeit the error bars of $\tau$ increase for larger $\phi$ due to the simultaneous reduction in $A$ we can exclude that a similarly strong variation as in $A$, which is up to 8 times, occurs for $\tau$. We rather find that $\tau$ depends only very weakly or is even independent on the FS angle. In particular the data for $F=32~\mu$J/cm$^2$ support this conclusion.

Here $A$ represents the density of photoexcited quasiparticles, which shows a momentum dependence strikingly similar to the gap function of a \textit{d}-wave superconductor. This can be understood by an indirect excitation process during which the order parameter is projected into the unoccupied electronic band structure. We now aim at an explanation of the processes active in the decay of that photoexcited state. Recalling that the pump-induced changes observed in Fig.~\ref{fig_trSCpeak}(a) occur within 50 meV around $E_\mathrm{F}$, i.e. close to the superconducting gap, we take QPs with energies on the order of the gap or smaller into account. We first consider QPs at $\phi=45^{\circ}$, i.e. at the node, where no gap is found, and illustrate QP relaxation in Fig.~\ref{fig_tauAndA}(c). At $T=0$ K, QPs with infinitesimally small energy can only scatter with other QP exactly at $\phi=45^{\circ}$, because at smaller $\phi$ the scattering partner cannot overcome the gap due to energy conservation. At higher $T$ QPs have larger energies and the scattering phase space is increased because the secondary QPs at other $k$ points can now overcome the gap near the nodal line. Far off the nodal line towards smaller $\phi$ this scattering channel is blocked for QP with energies about the gap size, as sketched in Fig.~\ref{fig_tauAndA}(d), process (1). Looking at our data we find that the amplitudes are actually larger far off the node and the decay is simply described for all momenta by a single exponential decay exhibiting constant $\tau$. Therefore, we find no indication of scattering towards the node. We conclude that albeit QPs just above the gap could gain energy through momentum redistribution (process (1), Fig.~\ref{fig_tauAndA}(d)) this channel is prohibited because the scattering partners required by momentum and energy conservation are not available. As a consequence QPs off the node become transiently stabilized, in agreement with the more sophisticated theoretical analysis of Refs.~\cite{Howell04,Gedik04}.

Having excluded intraband e-e scattering as a relaxation channel for the QP population above the gap, we face the question how to explain the observed 2.5~ps decay time. This time compares well to the one observed in earlier optical investigations \cite{Kaindl05,Demsar99,Liu08,Giannetti09}. We recall that on this very same time scale the recovery of the superconducting peak was observed here (Fig.~\ref{fig_trSCpeak}) and thus we can safely identify the decay of the QPs with recombination into Cooper pairs (process (2) in Fig.~\ref{fig_tauAndA}(d)). This requires coupling of a QP with momentum $\hbar k$ with one at $-\hbar k$ and a boson with twice the gap energy. The phenomenological Rothwarf-Taylor equations (RTE) \cite{Rothwarf67} are widely used to analyze such Cooper pair recombination \cite{Kabanov05,Kaindl05,Gedik04,Giannetti09,Kusar08}. At low $T$ where the photoexcited QP density, $n^*$, is considerably larger than the thermal one, $n_T$, i.e. $n^*\gg n_T$, the relaxation is governed by the probability to find a recombination partner leading to bimolecular recombination kinetics, as reported in \cite{Kaindl05} and explained in \cite{Kabanov05}. At higher $T<T_{c}$, in the so called boson bottleneck regime, an exponential decay is predicted and experimentally observed \cite{Liu08,Demsar99,Kusar05}. It arises from the competition between Cooper pair recombination emitting a boson with twice the gap energy and Cooper pair breaking by that boson. In this regime the relaxation is determined by the decay rate of these bosons and thus it is independent of the excitation density. Alternatively, the exponential decay observed at higher $T$, where $n_T>n^*$, has been also explained by means of the RTE not as due to a boson bottleneck but due to the recombination of the photoexcited QPs with the thermal ones \cite{Kaindl05}. The decay that we find at 30~K is well described by a single exponential function exhibiting momentum and pump fluence independent decay time. However, it cannot be explained considering only recombination with thermal QPs. Cooper pair formation requires the two QPs involved in the process to have opposite momenta and, as most of the thermally excited QPs are in the nodal region, this would imply that the photoexcited QPs scatter towards the node before they recombine with the thermal ones, contrary to our observation. Therefore only the existence of a boson bottleneck can explain all our results and we can conclude that Bi-2212 is in the strong bottleneck regime. This challenges the conclusion of previous optical studies \cite{Kaindl05,Gedik05} about the absence of a bottleneck, based on the observation of a bimolecular kinetics. However, such a dynamics can also be found in a strong bottleneck regime, as pointed out by an analytic solution of the RTE \cite{Kabanov05}.

Still, the intriguing momentum independence of the decay rates eludes explanation. The increase in the gap function for smaller $\phi$ should favor higher energy gain near the antinode increasing the decay rate. Also the momentum-dependence in the spin-fluctuation-mediated pairing interaction \cite{Dahm09} and in e-ph coupling \cite{bohnen03} suggests a variation of $\tau$ with $\phi$. However, in a \textit{d}-wave superconductor in the strong bottleneck regime, the interaction of $k$-dependent QP recombination, Cooper pair breaking and boson relaxation might give rise to a compensating effect, leading to the observed momentum independence of $\tau$. Here, a more sophisticated theoretical description beyond the RTE taking the symmetry of the order parameter into account would be necessary.

Finally we consider the second and faster component observed in the decay of $\Delta$$I(t)/I$ for $F=139~\mu$J/cm$^{2}$. Although our work aims particularly on the slower component, we note that the fast decay contribution is connected to scattering with QPs excited near the node at these higher pump fluences \cite{Howell04} and/or to a partial evaporation of the SC condensate \cite{Giannetti09,Kusar08}. However, further details are out of the scope of the current letter and will require additional studies as a function of the pump fluence and temperature.

In conclusion, we studied the momentum dependence of the transient population and decay times of photoexcited QPs in the high-$T_\mathrm{c}$ superconductor Bi-2212 by means of femtosecond trARPES. We observe a transient stabilization of the photoexcited QPs created by Cooper pair breaking, which is explained by blocking of e-e scattering away from the node. The decay of these QPs is well described by a single exponential with momentum and pump fluence independent decay time, which demonstrates that Bi-2212 is in the boson bottleneck regime.

R.C. acknowledges the Alexander von Humboldt Foundation. This work has been funded by the DFG through BO 1823/2-2.

\bibliography{Cortes_Bi2212_trARPES}

\begin{thebibliography}{18}%
\makeatletter
\providecommand \@ifxundefined [1]{%
 \@ifx{#1\undefined}
}%
\providecommand \@ifnum [1]{%
 \ifnum #1\expandafter \@firstoftwo
 \else \expandafter \@secondoftwo
 \fi
}%
\providecommand \@ifx [1]{%
 \ifx #1\expandafter \@firstoftwo
 \else \expandafter \@secondoftwo
 \fi
}%
\providecommand \natexlab [1]{#1}%
\providecommand \enquote  [1]{``#1''}%
\providecommand \bibnamefont  [1]{#1}%
\providecommand \bibfnamefont [1]{#1}%
\providecommand \citenamefont [1]{#1}%
\providecommand \href@noop [0]{\@secondoftwo}%
\providecommand \href [0]{\begingroup \@sanitize@url \@href}%
\providecommand \@href[1]{\@@startlink{#1}\@@href}%
\providecommand \@@href[1]{\endgroup#1\@@endlink}%
\providecommand \@sanitize@url [0]{\catcode `\\12\catcode `\$12\catcode
  `\&12\catcode `\#12\catcode `\^12\catcode `\_12\catcode `\%12\relax}%
\providecommand \@@startlink[1]{}%
\providecommand \@@endlink[0]{}%
\providecommand \url  [0]{\begingroup\@sanitize@url \@url }%
\providecommand \@url [1]{\endgroup\@href {#1}{\urlprefix }}%
\providecommand \urlprefix  [0]{URL }%
\providecommand \Eprint [0]{\href }%
\@ifxundefined \urlstyle {%
  \providecommand \doi  [0]{\begingroup \@sanitize@url \@doi}%
  \providecommand \@doi [1]{\endgroup \@@startlink {\doibase
  #1}doi:\discretionary {}{}{}#1\@@endlink }%
}{%
  \providecommand \doi  [0]{doi:\discretionary{}{}{}\begingroup
  \urlstyle{rm}\Url }%
}%
\providecommand \doibase [0]{http://dx.doi.org/}%
\providecommand \Doi [0]{\begingroup \@sanitize@url \@Doi }%
\providecommand \@Doi  [1]{\endgroup\@@startlink{\doibase#1}\@@Doi}%
\providecommand \@@Doi [1]{#1\@@endlink}%
\providecommand \selectlanguage [0]{\@gobble}%
\providecommand \bibinfo  [0]{\@secondoftwo}%
\providecommand \bibfield  [0]{\@secondoftwo}%
\providecommand \translation [1]{[#1]}%
\providecommand \BibitemOpen [0]{}%
\providecommand \bibitemStop [0]{}%
\providecommand \bibitemNoStop [0]{.\EOS\space}%
\providecommand \EOS [0]{\spacefactor3000\relax}%
\providecommand \BibitemShut  [1]{\csname bibitem#1\endcsname}%
\bibitem [{\citenamefont {Damascelli}\ \emph {et~al.}(2003)\citenamefont
  {Damascelli}, \citenamefont {Hussain},\ and\ \citenamefont
  {Shen}}]{Damascelli03}%
  \BibitemOpen
  \bibfield  {author} {\bibinfo {author} {\bibfnamefont {A.}~\bibnamefont
  {Damascelli}}, \bibinfo {author} {\bibfnamefont {Z.}~\bibnamefont {Hussain}},
  \ and\ \bibinfo {author} {\bibfnamefont {Z.-X.}\ \bibnamefont {Shen}},\ }\Doi
  {10.1103/RevModPhys.75.473} {\bibfield  {journal} {\bibinfo  {journal} {Rev.
  Mod. Phys.},\ }\textbf {\bibinfo {volume} {75}},\ \bibinfo {pages} {473}
  (\bibinfo {year} {2003})}\BibitemShut {NoStop}%
\bibitem [{\citenamefont {Kordyuk}\ \emph {et~al.}(2006)\citenamefont
  {Kordyuk}, \citenamefont {Borisenko}, \citenamefont {Zabolotnyy},
  \citenamefont {Geck}, \citenamefont {Knupfer}, \citenamefont {Fink},
  \citenamefont {B\"uchner}, \citenamefont {Lin}, \citenamefont {Keimer},
  \citenamefont {Berger}, \citenamefont {Pan}, \citenamefont {Komiya},\ and\
  \citenamefont {Ando}}]{Kordyuk06}%
  \BibitemOpen
  \bibfield  {author} {\bibinfo {author} {\bibfnamefont {A.~A.}\ \bibnamefont
  {Kordyuk}}, \bibinfo {author} {\bibfnamefont {S.~V.}\ \bibnamefont
  {Borisenko}}, \bibinfo {author} {\bibfnamefont {V.~B.}\ \bibnamefont
  {Zabolotnyy}}, \bibinfo {author} {\bibfnamefont {J.}~\bibnamefont {Geck}},
  \bibinfo {author} {\bibfnamefont {M.}~\bibnamefont {Knupfer}}, \bibinfo
  {author} {\bibfnamefont {J.}~\bibnamefont {Fink}}, \bibinfo {author}
  {\bibfnamefont {B.}~\bibnamefont {B\"uchner}}, \bibinfo {author}
  {\bibfnamefont {C.~T.}\ \bibnamefont {Lin}}, \bibinfo {author} {\bibfnamefont
  {B.}~\bibnamefont {Keimer}}, \bibinfo {author} {\bibfnamefont
  {H.}~\bibnamefont {Berger}}, \bibinfo {author} {\bibfnamefont {A.~V.}\
  \bibnamefont {Pan}}, \bibinfo {author} {\bibfnamefont {S.}~\bibnamefont
  {Komiya}}, \ and\ \bibinfo {author} {\bibfnamefont {Y.}~\bibnamefont
  {Ando}},\ }\Doi {10.1103/PhysRevLett.97.017002} {\bibfield  {journal}
  {\bibinfo  {journal} {Phys. Rev. Lett.},\ }\textbf {\bibinfo {volume} {97}},\
  \bibinfo {pages} {017002} (\bibinfo {year} {2006})}\BibitemShut {NoStop}%
\bibitem [{\citenamefont {Demsar}\ \emph {et~al.}(1999)\citenamefont {Demsar},
  \citenamefont {Podobnik}, \citenamefont {Kabanov}, \citenamefont {Wolf},\
  and\ \citenamefont {Mihailovic}}]{Demsar99}%
  \BibitemOpen
  \bibfield  {author} {\bibinfo {author} {\bibfnamefont {J.}~\bibnamefont
  {Demsar}}, \bibinfo {author} {\bibfnamefont {B.}~\bibnamefont {Podobnik}},
  \bibinfo {author} {\bibfnamefont {V.~V.}\ \bibnamefont {Kabanov}}, \bibinfo
  {author} {\bibfnamefont {T.}~\bibnamefont {Wolf}}, \ and\ \bibinfo {author}
  {\bibfnamefont {D.}~\bibnamefont {Mihailovic}},\ }\href@noop {} {\bibfield
  {journal} {\bibinfo  {journal} {Phys. Rev. Lett.},\ }\textbf {\bibinfo
  {volume} {82}},\ \bibinfo {pages} {4918} (\bibinfo {year}
  {1999})}\BibitemShut {NoStop}%
\bibitem [{\citenamefont {Gedik}\ \emph {et~al.}(2004)\citenamefont {Gedik},
  \citenamefont {Blake}, \citenamefont {Spitzer}, \citenamefont {Orenstein},
  \citenamefont {Liang}, \citenamefont {Bonn},\ and\ \citenamefont
  {Hardy}}]{Gedik04}%
  \BibitemOpen
  \bibfield  {author} {\bibinfo {author} {\bibfnamefont {N.}~\bibnamefont
  {Gedik}}, \bibinfo {author} {\bibfnamefont {P.}~\bibnamefont {Blake}},
  \bibinfo {author} {\bibfnamefont {R.~C.}\ \bibnamefont {Spitzer}}, \bibinfo
  {author} {\bibfnamefont {J.}~\bibnamefont {Orenstein}}, \bibinfo {author}
  {\bibfnamefont {R.~X.}\ \bibnamefont {Liang}}, \bibinfo {author}
  {\bibfnamefont {D.~A.}\ \bibnamefont {Bonn}}, \ and\ \bibinfo {author}
  {\bibfnamefont {W.~N.}\ \bibnamefont {Hardy}},\ }\Doi
  {10.1103/PhysRevB.70.014504} {\bibfield  {journal} {\bibinfo  {journal}
  {Phys. Rev. B},\ }\textbf {\bibinfo {volume} {70}},\ \bibinfo {pages}
  {014504} (\bibinfo {year} {2004})}\BibitemShut {NoStop}%
\bibitem [{\citenamefont {Kusar}\ \emph {et~al.}(2005)\citenamefont {Kusar},
  \citenamefont {Demsar}, \citenamefont {Mihailovic},\ and\ \citenamefont
  {Sugai}}]{Kusar05}%
  \BibitemOpen
  \bibfield  {author} {\bibinfo {author} {\bibfnamefont {P.}~\bibnamefont
  {Kusar}}, \bibinfo {author} {\bibfnamefont {J.}~\bibnamefont {Demsar}},
  \bibinfo {author} {\bibfnamefont {D.}~\bibnamefont {Mihailovic}}, \ and\
  \bibinfo {author} {\bibfnamefont {S.}~\bibnamefont {Sugai}},\ }\Doi
  {10.1103/PhysRevB.72.014544} {\bibfield  {journal} {\bibinfo  {journal}
  {Phys. Rev. B},\ }\textbf {\bibinfo {volume} {72}},\ \bibinfo {pages}
  {014544} (\bibinfo {year} {2005})}\BibitemShut {NoStop}%
\bibitem [{\citenamefont {Kaindl}\ \emph {et~al.}(2005)\citenamefont {Kaindl},
  \citenamefont {Carnahan}, \citenamefont {Chemla}, \citenamefont {Oh},\ and\
  \citenamefont {Eckstein}}]{Kaindl05}%
  \BibitemOpen
  \bibfield  {author} {\bibinfo {author} {\bibfnamefont {R.~A.}\ \bibnamefont
  {Kaindl}}, \bibinfo {author} {\bibfnamefont {M.~A.}\ \bibnamefont
  {Carnahan}}, \bibinfo {author} {\bibfnamefont {D.~S.}\ \bibnamefont
  {Chemla}}, \bibinfo {author} {\bibfnamefont {S.}~\bibnamefont {Oh}}, \ and\
  \bibinfo {author} {\bibfnamefont {J.~N.}\ \bibnamefont {Eckstein}},\ }\Doi
  {10.1103/PhysRevB.72.060510} {\bibfield  {journal} {\bibinfo  {journal}
  {Phys. Rev. B},\ }\textbf {\bibinfo {volume} {72}},\ \bibinfo {pages}
  {060510(R)} (\bibinfo {year} {2005})}\BibitemShut {NoStop}%
\bibitem [{\citenamefont {Gedik}\ \emph {et~al.}(2005)\citenamefont {Gedik},
  \citenamefont {Langner}, \citenamefont {Orenstein}, \citenamefont {Ono},
  \citenamefont {Abe},\ and\ \citenamefont {Ando}}]{Gedik05}%
  \BibitemOpen
  \bibfield  {author} {\bibinfo {author} {\bibfnamefont {N.}~\bibnamefont
  {Gedik}}, \bibinfo {author} {\bibfnamefont {M.}~\bibnamefont {Langner}},
  \bibinfo {author} {\bibfnamefont {J.}~\bibnamefont {Orenstein}}, \bibinfo
  {author} {\bibfnamefont {S.}~\bibnamefont {Ono}}, \bibinfo {author}
  {\bibfnamefont {Y.}~\bibnamefont {Abe}}, \ and\ \bibinfo {author}
  {\bibfnamefont {Y.}~\bibnamefont {Ando}},\ }\Doi
  {10.1103/PhysRevLett.95.117005} {\bibfield  {journal} {\bibinfo  {journal}
  {Phys. Rev. Lett.},\ }\textbf {\bibinfo {volume} {95}},\ \bibinfo {pages}
  {117005} (\bibinfo {year} {2005})}\BibitemShut {NoStop}%
\bibitem [{\citenamefont {Kusar}\ \emph {et~al.}(2008)\citenamefont {Kusar},
  \citenamefont {Kabanov}, \citenamefont {Sugai}, \citenamefont {Demsar},
  \citenamefont {Mertelj},\ and\ \citenamefont {Mihailovic}}]{Kusar08}%
  \BibitemOpen
  \bibfield  {author} {\bibinfo {author} {\bibfnamefont {P.}~\bibnamefont
  {Kusar}}, \bibinfo {author} {\bibfnamefont {V.~V.}\ \bibnamefont {Kabanov}},
  \bibinfo {author} {\bibfnamefont {S.}~\bibnamefont {Sugai}}, \bibinfo
  {author} {\bibfnamefont {J.}~\bibnamefont {Demsar}}, \bibinfo {author}
  {\bibfnamefont {T.}~\bibnamefont {Mertelj}}, \ and\ \bibinfo {author}
  {\bibfnamefont {D.}~\bibnamefont {Mihailovic}},\ }\Doi
  {10.1103/PhysRevLett.101.227001} {\bibfield  {journal} {\bibinfo  {journal}
  {Phys. Rev. Lett.},\ }\textbf {\bibinfo {volume} {101}},\ \bibinfo {pages}
  {227001} (\bibinfo {year} {2008})}\BibitemShut {NoStop}%
\bibitem [{\citenamefont {Liu}\ \emph {et~al.}(2008)\citenamefont {Liu},
  \citenamefont {Toda}, \citenamefont {Shimatake}, \citenamefont {Momono},
  \citenamefont {Oda},\ and\ \citenamefont {Ido}}]{Liu08}%
  \BibitemOpen
  \bibfield  {author} {\bibinfo {author} {\bibfnamefont {Y.~H.}\ \bibnamefont
  {Liu}}, \bibinfo {author} {\bibfnamefont {Y.}~\bibnamefont {Toda}}, \bibinfo
  {author} {\bibfnamefont {K.}~\bibnamefont {Shimatake}}, \bibinfo {author}
  {\bibfnamefont {N.}~\bibnamefont {Momono}}, \bibinfo {author} {\bibfnamefont
  {M.}~\bibnamefont {Oda}}, \ and\ \bibinfo {author} {\bibfnamefont
  {M.}~\bibnamefont {Ido}},\ }\Doi {10.1103/PhysRevLett.101.137003} {\bibfield
  {journal} {\bibinfo  {journal} {Phys. Rev. Lett.},\ }\textbf {\bibinfo
  {volume} {101}},\ \bibinfo {pages} {137003} (\bibinfo {year}
  {2008})}\BibitemShut {NoStop}%
\bibitem [{\citenamefont {Giannetti}\ \emph {et~al.}(2009)\citenamefont
  {Giannetti}, \citenamefont {Coslovich}, \citenamefont {Cilento},
  \citenamefont {Ferrini}, \citenamefont {Eisaki}, \citenamefont {Kaneko},
  \citenamefont {Greven},\ and\ \citenamefont {Parmigiani}}]{Giannetti09}%
  \BibitemOpen
  \bibfield  {author} {\bibinfo {author} {\bibfnamefont {C.}~\bibnamefont
  {Giannetti}}, \bibinfo {author} {\bibfnamefont {G.}~\bibnamefont
  {Coslovich}}, \bibinfo {author} {\bibfnamefont {F.}~\bibnamefont {Cilento}},
  \bibinfo {author} {\bibfnamefont {G.}~\bibnamefont {Ferrini}}, \bibinfo
  {author} {\bibfnamefont {H.}~\bibnamefont {Eisaki}}, \bibinfo {author}
  {\bibfnamefont {N.}~\bibnamefont {Kaneko}}, \bibinfo {author} {\bibfnamefont
  {M.}~\bibnamefont {Greven}}, \ and\ \bibinfo {author} {\bibfnamefont
  {F.}~\bibnamefont {Parmigiani}},\ }\Doi {10.1103/PhysRevB.79.224502}
  {\bibfield  {journal} {\bibinfo  {journal} {Phys. Rev. B},\ }\textbf
  {\bibinfo {volume} {79}},\ \bibinfo {pages} {224502} (\bibinfo {year}
  {2009})}\BibitemShut {NoStop}%
\bibitem [{\citenamefont {Kabanov}\ \emph {et~al.}(2005)\citenamefont
  {Kabanov}, \citenamefont {Demsar},\ and\ \citenamefont
  {Mihailovic}}]{Kabanov05}%
  \BibitemOpen
  \bibfield  {author} {\bibinfo {author} {\bibfnamefont {V.~V.}\ \bibnamefont
  {Kabanov}}, \bibinfo {author} {\bibfnamefont {J.}~\bibnamefont {Demsar}}, \
  and\ \bibinfo {author} {\bibfnamefont {D.}~\bibnamefont {Mihailovic}},\ }\Doi
  {10.1103/PhysRevLett.95.147002} {\bibfield  {journal} {\bibinfo  {journal}
  {Phys. Rev. Lett.},\ }\textbf {\bibinfo {volume} {95}},\ \bibinfo {pages}
  {147002} (\bibinfo {year} {2005})}\BibitemShut {NoStop}%
\bibitem [{\citenamefont {Howell}\ \emph {et~al.}(2004)\citenamefont {Howell},
  \citenamefont {Rosch},\ and\ \citenamefont {Hirschfeld}}]{Howell04}%
  \BibitemOpen
  \bibfield  {author} {\bibinfo {author} {\bibfnamefont {P.~C.}\ \bibnamefont
  {Howell}}, \bibinfo {author} {\bibfnamefont {A.}~\bibnamefont {Rosch}}, \
  and\ \bibinfo {author} {\bibfnamefont {P.~J.}\ \bibnamefont {Hirschfeld}},\
  }\Doi {10.1103/PhysRevLett.92.037003} {\bibfield  {journal} {\bibinfo
  {journal} {Phys. Rev. Lett.},\ }\textbf {\bibinfo {volume} {92}},\ \bibinfo
  {pages} {037003} (\bibinfo {year} {2004})}\BibitemShut {NoStop}%
\bibitem [{\citenamefont {Perfetti}\ \emph {et~al.}(2007)\citenamefont
  {Perfetti}, \citenamefont {Loukakos}, \citenamefont {Lisowski}, \citenamefont
  {Bovensiepen}, \citenamefont {Eisaki},\ and\ \citenamefont
  {Wolf}}]{Perfetti07}%
  \BibitemOpen
  \bibfield  {author} {\bibinfo {author} {\bibfnamefont {L.}~\bibnamefont
  {Perfetti}}, \bibinfo {author} {\bibfnamefont {P.~A.}\ \bibnamefont
  {Loukakos}}, \bibinfo {author} {\bibfnamefont {M.}~\bibnamefont {Lisowski}},
  \bibinfo {author} {\bibfnamefont {U.}~\bibnamefont {Bovensiepen}}, \bibinfo
  {author} {\bibfnamefont {H.}~\bibnamefont {Eisaki}}, \ and\ \bibinfo {author}
  {\bibfnamefont {M.}~\bibnamefont {Wolf}},\ }\Doi
  {10.1103/PhysRevLett.99.197001} {\bibfield  {journal} {\bibinfo  {journal}
  {Phys. Rev. Lett.},\ }\textbf {\bibinfo {volume} {99}},\ \bibinfo {pages}
  {197001} (\bibinfo {year} {2007})}\BibitemShut {NoStop}%
\bibitem [{\citenamefont {Schmitt}\ \emph {et~al.}(2008)\citenamefont
  {Schmitt}, \citenamefont {Kirchmann}, \citenamefont {Bovensiepen},
  \citenamefont {Moore}, \citenamefont {Rettig}, \citenamefont {Krenz},
  \citenamefont {Chu}, \citenamefont {Ru}, \citenamefont {Perfetti},
  \citenamefont {Lu}, \citenamefont {Wolf}, \citenamefont {Fisher},\ and\
  \citenamefont {Shen}}]{Schmitt08}%
  \BibitemOpen
  \bibfield  {author} {\bibinfo {author} {\bibfnamefont {F.}~\bibnamefont
  {Schmitt}}, \bibinfo {author} {\bibfnamefont {P.~S.}\ \bibnamefont
  {Kirchmann}}, \bibinfo {author} {\bibfnamefont {U.}~\bibnamefont
  {Bovensiepen}}, \bibinfo {author} {\bibfnamefont {R.~G.}\ \bibnamefont
  {Moore}}, \bibinfo {author} {\bibfnamefont {L.}~\bibnamefont {Rettig}},
  \bibinfo {author} {\bibfnamefont {M.}~\bibnamefont {Krenz}}, \bibinfo
  {author} {\bibfnamefont {J.~H.}\ \bibnamefont {Chu}}, \bibinfo {author}
  {\bibfnamefont {N.}~\bibnamefont {Ru}}, \bibinfo {author} {\bibfnamefont
  {L.}~\bibnamefont {Perfetti}}, \bibinfo {author} {\bibfnamefont {D.~H.}\
  \bibnamefont {Lu}}, \bibinfo {author} {\bibfnamefont {M.}~\bibnamefont
  {Wolf}}, \bibinfo {author} {\bibfnamefont {I.~R.}\ \bibnamefont {Fisher}}, \
  and\ \bibinfo {author} {\bibfnamefont {Z.~X.}\ \bibnamefont {Shen}},\ }\Doi
  {10.1126/science.1160778} {\bibfield  {journal} {\bibinfo  {journal}
  {Science},\ }\textbf {\bibinfo {volume} {321}},\ \bibinfo {pages} {1649}
  (\bibinfo {year} {2008})}\BibitemShut {NoStop}%
\bibitem [{\citenamefont {Schoenlein}\ \emph {et~al.}(1987)\citenamefont
  {Schoenlein}, \citenamefont {Lin}, \citenamefont {Fujimoto},\ and\
  \citenamefont {Eesley}}]{Schoenlein87}%
  \BibitemOpen
  \bibfield  {author} {\bibinfo {author} {\bibfnamefont {R.~W.}\ \bibnamefont
  {Schoenlein}}, \bibinfo {author} {\bibfnamefont {W.~Z.}\ \bibnamefont {Lin}},
  \bibinfo {author} {\bibfnamefont {J.~G.}\ \bibnamefont {Fujimoto}}, \ and\
  \bibinfo {author} {\bibfnamefont {G.~L.}\ \bibnamefont {Eesley}},\ }\Doi
  {10.1103/PhysRevLett.58.1680} {\bibfield  {journal} {\bibinfo  {journal}
  {Phys. Rev. Lett.},\ }\textbf {\bibinfo {volume} {58}},\ \bibinfo {pages}
  {1680} (\bibinfo {year} {1987})}\BibitemShut {NoStop}%
\bibitem [{\citenamefont {Rothwarf}\ and\ \citenamefont
  {Taylor}(1967)}]{Rothwarf67}%
  \BibitemOpen
  \bibfield  {author} {\bibinfo {author} {\bibfnamefont {A.}~\bibnamefont
  {Rothwarf}}\ and\ \bibinfo {author} {\bibfnamefont {B.~N.}\ \bibnamefont
  {Taylor}},\ }\Doi {10.1103/PhysRevLett.19.27} {\bibfield  {journal} {\bibinfo
   {journal} {Phys. Rev. Lett.},\ }\textbf {\bibinfo {volume} {19}},\ \bibinfo
  {pages} {27} (\bibinfo {year} {1967})}\BibitemShut {NoStop}%
\bibitem [{\citenamefont {Dahm}\ \emph {et~al.}(2009)\citenamefont {Dahm},
  \citenamefont {Hinkov}, \citenamefont {Borisenko}, \citenamefont {Kordyuk},
  \citenamefont {Zabolotnyy}, \citenamefont {Fink}, \citenamefont {Buechner},
  \citenamefont {Scalapino}, \citenamefont {Hanke},\ and\ \citenamefont
  {Keimer}}]{Dahm09}%
  \BibitemOpen
  \bibfield  {author} {\bibinfo {author} {\bibfnamefont {T.}~\bibnamefont
  {Dahm}}, \bibinfo {author} {\bibfnamefont {V.}~\bibnamefont {Hinkov}},
  \bibinfo {author} {\bibfnamefont {S.~V.}\ \bibnamefont {Borisenko}}, \bibinfo
  {author} {\bibfnamefont {A.~A.}\ \bibnamefont {Kordyuk}}, \bibinfo {author}
  {\bibfnamefont {V.~B.}\ \bibnamefont {Zabolotnyy}}, \bibinfo {author}
  {\bibfnamefont {J.}~\bibnamefont {Fink}}, \bibinfo {author} {\bibfnamefont
  {B.}~\bibnamefont {Buechner}}, \bibinfo {author} {\bibfnamefont {D.~J.}\
  \bibnamefont {Scalapino}}, \bibinfo {author} {\bibfnamefont {W.}~\bibnamefont
  {Hanke}}, \ and\ \bibinfo {author} {\bibfnamefont {B.}~\bibnamefont
  {Keimer}},\ }\Doi {10.1038/NPHYS1180} {\bibfield  {journal} {\bibinfo
  {journal} {Nature Physics},\ }\textbf {\bibinfo {volume} {5}},\ \bibinfo
  {pages} {217} (\bibinfo {year} {2009})}\BibitemShut {NoStop}%
\bibitem [{\citenamefont {Bohnen}\ \emph {et~al.}(2003)\citenamefont {Bohnen},
  \citenamefont {Heid},\ and\ \citenamefont {Krauss}}]{bohnen03}%
  \BibitemOpen
  \bibfield  {author} {\bibinfo {author} {\bibfnamefont {K.-P.}\ \bibnamefont
  {Bohnen}}, \bibinfo {author} {\bibfnamefont {R.}~\bibnamefont {Heid}}, \ and\
  \bibinfo {author} {\bibfnamefont {M.}~\bibnamefont {Krauss}},\ }\href@noop {}
  {\bibfield  {journal} {\bibinfo  {journal} {Europhys. Lett.},\ }\textbf
  {\bibinfo {volume} {64}},\ \bibinfo {pages} {104} (\bibinfo {year}
  {2003})}\BibitemShut {NoStop}%
\end{thebibliography}%

\end{document}